\documentclass[aps,prd,longbibliography,twocolumn,showpacs,superscriptaddress,floatfix,10pt]{revtex4-1}  

\usepackage{hyperref}
\usepackage{graphicx}
\usepackage{dcolumn}
\usepackage{bm}
\usepackage{amsmath}
\usepackage{amsthm}
\usepackage[dvipsnames]{xcolor}
\usepackage{lineno}
\usepackage{soul}
\usepackage{placeins} 
\graphicspath{{figures/}}

\begin{document}

\title{First Measurement of Charged Current Muon Neutrino-Induced \texorpdfstring{$K^+$}{TEXT} Production on Argon using the MicroBooNE Detector}

\newcommand{\ANL}{Argonne National Laboratory (ANL), Lemont, IL, 60439, USA}
\newcommand{\Bern}{Universit{\"a}t Bern, Bern CH-3012, Switzerland}
\newcommand{\BNL}{Brookhaven National Laboratory (BNL), Upton, NY, 11973, USA}
\newcommand{\UCSB}{University of California, Santa Barbara, CA, 93106, USA}
\newcommand{\Cambridge}{University of Cambridge, Cambridge CB3 0HE, United Kingdom}
\newcommand{\CIEMAT}{Centro de Investigaciones Energ\'{e}ticas, Medioambientales y Tecnol\'{o}gicas (CIEMAT), Madrid E-28040, Spain}
\newcommand{\Chicago}{University of Chicago, Chicago, IL, 60637, USA}
\newcommand{\Cincinnati}{University of Cincinnati, Cincinnati, OH, 45221, USA}
\newcommand{\CSU}{Colorado State University, Fort Collins, CO, 80523, USA}
\newcommand{\Columbia}{Columbia University, New York, NY, 10027, USA}
\newcommand{\Edinburgh}{University of Edinburgh, Edinburgh EH9 3FD, United Kingdom}
\newcommand{\FNAL}{Fermi National Accelerator Laboratory (FNAL), Batavia, IL 60510, USA}
\newcommand{\Granada}{Universidad de Granada, Granada E-18071, Spain}
\newcommand{\IIT}{Illinois Institute of Technology (IIT), Chicago, IL 60616, USA}
\newcommand{\ICL}{Imperial College London, London SW7 2AZ, United Kingdom}
\newcommand{\Indiana}{Indiana University, Bloomington, IN 47405, USA}
\newcommand{\Kansas}{The University of Kansas, Lawrence, KS, 66045, USA}
\newcommand{\KSU}{Kansas State University (KSU), Manhattan, KS, 66506, USA}
\newcommand{\Lancaster}{Lancaster University, Lancaster LA1 4YW, United Kingdom}
\newcommand{\LANL}{Los Alamos National Laboratory (LANL), Los Alamos, NM, 87545, USA}
\newcommand{\Louisiana}{Louisiana State University, Baton Rouge, LA, 70803, USA}
\newcommand{\Manchester}{The University of Manchester, Manchester M13 9PL, United Kingdom}
\newcommand{\MIT}{Massachusetts Institute of Technology (MIT), Cambridge, MA, 02139, USA}
\newcommand{\Michigan}{University of Michigan, Ann Arbor, MI, 48109, USA}
\newcommand{\MSU}{Michigan State University, East Lansing, MI 48824, USA}
\newcommand{\Minnesota}{University of Minnesota, Minneapolis, MN, 55455, USA}
\newcommand{\Nankai}{Nankai University, Nankai District, Tianjin 300071, China}
\newcommand{\NMSU}{New Mexico State University (NMSU), Las Cruces, NM, 88003, USA}
\newcommand{\Oxford}{University of Oxford, Oxford OX1 3RH, United Kingdom}
\newcommand{\Pitt}{University of Pittsburgh, Pittsburgh, PA, 15260, USA}
\newcommand{\QMUL}{Queen Mary University of London, London E1 4NS, United Kingdom}
\newcommand{\Rutgers}{Rutgers University, Piscataway, NJ, 08854, USA}
\newcommand{\SLAC}{SLAC National Accelerator Laboratory, Menlo Park, CA, 94025, USA}
\newcommand{\SDSMT}{South Dakota School of Mines and Technology (SDSMT), Rapid City, SD, 57701, USA}
\newcommand{\Maine}{University of Southern Maine, Portland, ME, 04104, USA}
\newcommand{\Syracuse}{Syracuse University, Syracuse, NY, 13244, USA}
\newcommand{\TelAviv}{Tel Aviv University, Tel Aviv, Israel, 69978}
\newcommand{\UTA}{University of Texas, Arlington, TX, 76019, USA}
\newcommand{\Tufts}{Tufts University, Medford, MA, 02155, USA}
\newcommand{\VTech}{Center for Neutrino Physics, Virginia Tech, Blacksburg, VA, 24061, USA}
\newcommand{\Warwick}{University of Warwick, Coventry CV4 7AL, United Kingdom}

\affiliation{\ANL}
\affiliation{\Bern}
\affiliation{\BNL}
\affiliation{\UCSB}
\affiliation{\Cambridge}
\affiliation{\CIEMAT}
\affiliation{\Chicago}
\affiliation{\Cincinnati}
\affiliation{\CSU}
\affiliation{\Columbia}
\affiliation{\Edinburgh}
\affiliation{\FNAL}
\affiliation{\Granada}
\affiliation{\IIT}
\affiliation{\ICL}
\affiliation{\Indiana}
\affiliation{\Kansas}
\affiliation{\KSU}
\affiliation{\Lancaster}
\affiliation{\LANL}
\affiliation{\Louisiana}
\affiliation{\Manchester}
\affiliation{\MIT}
\affiliation{\Michigan}
\affiliation{\MSU}
\affiliation{\Minnesota}
\affiliation{\Nankai}
\affiliation{\NMSU}
\affiliation{\Oxford}
\affiliation{\Pitt}
\affiliation{\QMUL}
\affiliation{\Rutgers}
\affiliation{\SLAC}
\affiliation{\SDSMT}
\affiliation{\Maine}
\affiliation{\Syracuse}
\affiliation{\TelAviv}
\affiliation{\UTA}
\affiliation{\Tufts}
\affiliation{\VTech}
\affiliation{\Warwick}

\author{P.~Abratenko} \affiliation{\Tufts}
\author{D.~Andrade~Aldana} \affiliation{\IIT}
\author{L.~Arellano} \affiliation{\Manchester}
\author{J.~Asaadi} \affiliation{\UTA}
\author{A.~Ashkenazi}\affiliation{\TelAviv}
\author{S.~Balasubramanian}\affiliation{\FNAL}
\author{B.~Baller} \affiliation{\FNAL}
\author{A.~Barnard} \affiliation{\Oxford}
\author{G.~Barr} \affiliation{\Oxford}
\author{D.~Barrow} \affiliation{\Oxford}
% ...

\collaboration{The MicroBooNE Collaboration}
\thanks{microboone\_info@fnal.gov}\noaffiliation

\begin{abstract}

The MicroBooNE experiment is an 85 tonne active mass liquid argon time projection chamber neutrino detector exposed to the on-axis Booster Neutrino Beam (BNB) at Fermilab. One of MicroBooNE's physics goals is the precise measurement of neutrino interactions on argon in the 1 GeV energy regime. Building on the capabilities of the MicroBooNE detector, this analysis identifies $K^{+}$ mesons, a key signature for the study of strange particle production in neutrino interactions. This measurement is furthermore valuable for background estimation for future nucleon decay searches and for improved reconstruction and particle identification capabilities in experiments such as the Deep Underground Neutrino Experiment (DUNE). In this letter, we present the first-ever measurement of a flux-integrated cross section for charged-current muon neutrino induced $K^{+}$ production on argon nuclei, determined to be 7.93 $\pm$ 3.22 (stat.) $\pm$ 2.83 (syst.) $\times~10^{-42}\;$ cm$^2$/nucleon based on an analysis of 6.88$\times10^{20}$ protons on target. This result was found to be consistent with model predictions from different neutrino event generators within the reported uncertainties.

\end{abstract}

\maketitle
Within the Standard Model (SM), the stability of protons is a consequence of the global symmetry that conserves baryon number. However, various Grand Unified Theories (GUTs) predicting baryon number violation through nucleon decay suggest different decay channels, with a preference for $p \rightarrow e^{+} \pi^{0}$, with lifetimes ranging from $10^{34}$ to $10^{39}$ years \cite{PhysRevLett.32.438}. Similarly, the Supersymmetric Grand Unified Theory (SUSY GUT) favors protons decaying via $p \rightarrow \nu K^{+}$, with lifetimes of $10^{32}$ to $10^{35}$ years \cite{Hisano:1992jj}.

Upcoming large long-baseline neutrino oscillation experiments, such as the Deep Underground Neutrino Experiment (DUNE) \cite{dunetdr} and Hyper-Kamiokande \cite{protocollaboration2018hyperkamiokandedesignreport}, have the sensitivity to search for proton decay in the range of $10^{32}$ to $10^{39}$ years. Cherenkov detector experiments like Super-Kamiokande \cite{PhysRevD.90.072005, PhysRevD.95.012004, PhysRevD.102.112011} and Hyper-Kamiokande are particularly suited for $p \rightarrow e^{+} \pi^{0}$ decay searches. However, for water Cherenkov detectors, the $p \rightarrow \nu K^{+}$ decay channel needs to be indirectly detected since the momentum of the produced $K^{+}$ (339 MeV/c) is below the Cherenkov detection threshold in water (560 MeV/c). Furthermore, backgrounds from atmospheric neutrino interactions, such as the $ \nu_\mu n \rightarrow \mu^- K^+ n $ and $ \nu p \rightarrow \nu K^+ \Lambda $ channels, are challenging to distinguish in water Cherenkov detectors.

In contrast, liquid argon time projection chambers (LArTPCs) can directly measure 
kaons that range out and decay by analyzing their unique energy loss ($dE/dx$) profile. These profiles, along with topological signatures such as distinctive kinks from kaon decay and the high ionization Bragg peak, allow an enhanced kaon identification. All of these topological and calorimetric features are easily measured using the millimiter-precision imaging capabilities of a LArTPC. 

The first measurements of charged-current (CC) and neutral-current (NC) kaon production, reported with the Argonne National Laboratory \cite{PhysRevLett.33.1446} and GARGAMELLE bubble chambers \cite{osti_4109186} in 1974 and 1975, observed $\mathcal{O}$(10) events. Later, in the 1980's, the Brookhaven National Laboratory bubble chamber reported $\mathcal{O}$(10) CC and NC events with strange particle production in neutrino-hydrogen and neutrino-deuterium interactions \cite{Jensen:1964zz, PhysRevD.24.2779}, and the Fermilab bubble chamber reported $\mathcal{O}$(10) events and a measured cross section for CC neutrino interactions $\nu_\mu + n \xrightarrow{} \mu^- + \Lambda + K^+$ on a deuterium target for energy ranges between 10 GeV and 250 GeV \cite{PhysRevD.28.2129}. Most recently, the MINERvA experiment measured $K^+$ production in NC and CC modes on a hydrocarbon target, reporting $\mathcal{O}$(100) and $\mathcal{O}$(1000) events respectively \cite{PhysRevD.94.012002, PhysRevLett.119.011802, Marshall:2016mar}. In addition, MINERvA reported the first evidence of coherent kaon production \cite{PhysRevLett.117.061802, Marshall:2016mar}.

This letter presents the first measurement of the cross section for $K^{+}$ production in neutrino interactions on argon nuclei. The signature of the signal events is a CC $\nu_\mu$ interaction in the presence of a $K^{+}$ in the final state with momentum $230~\text{MeV/c} < p_{K^+} < 2900~\text{MeV/c}$. The $K^{+}$ should decay either to $ \mu^+ \nu_\mu $ or $ \pi^+ \pi^0 $ which have branching ratios of 63.56\% and 20.67\%, respectively \cite{PhysRevD.110.030001}. 
This measurement uses MicroBooNE datasets collected between 2016 and 2018 from the on-axis BNB, equivalent to 6.88$\times10^{20}$ protons on target (POT). 

As the first quantitative measurement of $K^{+}$ production on argon nuclei, this analysis will impact nuclear, neutrino, and proton decay research on two fronts. First, it provides background constraints in future $p \rightarrow \nu K^{+}$ nucleon decay experiments.
Second, this measurement provides input to improve neutrino generator models. In the case of bound proton decay, the $K^{+}$ is generated within the nucleus and may interact with other nucleons as it escapes. These interactions can decrease the detected momentum of the $K^{+}$ in the detector and may also eject nucleons from the nucleus.

\begin{figure}[htbp]
\centering
\includegraphics[width=0.95\linewidth]{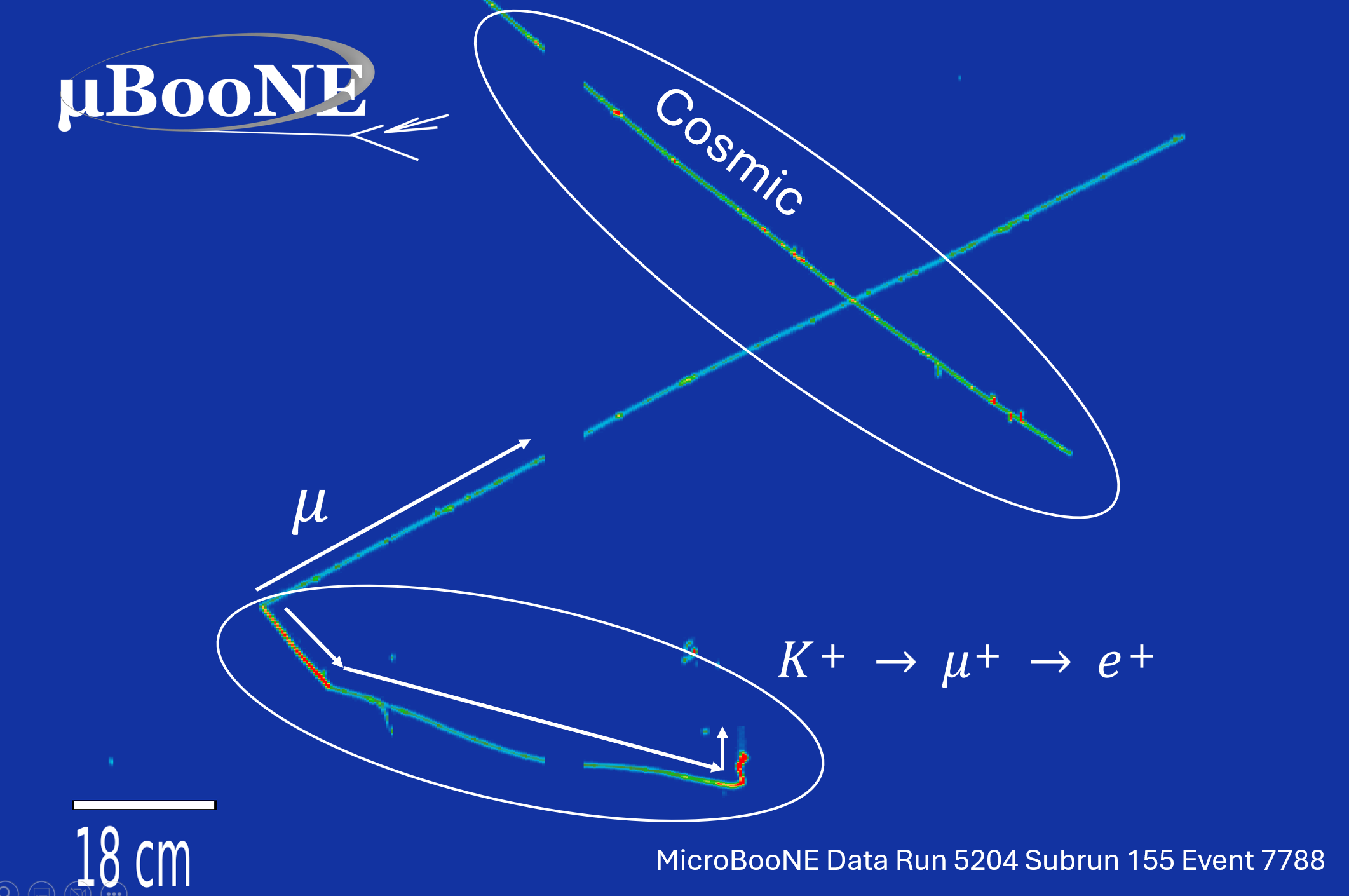}
\caption{A $\nu_{\mu} + \text{Ar} \rightarrow \mu^{-} + K^{+}$ interaction candidate observed in MicroBooNE data recorded by the collection plane. The color scale represents the intensity of the ionization charge collected on the TPC wires (green: low intensity, red: high intensity). The x and y-axes represent the wire position in the beam direction and the distance in the drift direction, respectively. The gap shown on each track is due to a region in the detector with no active wires.} 
\label{fig:evd}
\end{figure}

Charged-current neutrino-induced $K^+$ production occurs through three interaction modes: (1) \textit{single kaon production}, where a CC interaction produces a single kaon with a production energy threshold of  $E_{\nu_\mu}$ $\geq$ 0.79 GeV, and strangeness is not conserved ($|\Delta S|=1$); (2) \textit{associated kaon production}, where both a kaon and a hyperon are produced, with a threshold of $E_{\nu_\mu}$ $\geq$ 1.2 GeV, and strangeness is conserved; and (3) \textit{coherent production}, where only one kaon is produced in the final state, and the target nucleus remains intact after the neutrino interaction. Figure \ref{fig:evd} shows an event candidate of CC $\nu_\mu$ kaon production recorded in MicroBooNE data. The kaon decays at rest into a 388 MeV energy muon, which subsequently decays into a Michel electron.

The MicroBooNE detector \cite{Fleming:2012gvl, MicroBooNE:2016pwy} consists of an 85 tonne LArTPC and a photon detection system comprised of 32 photomultiplier tubes (PMTs). The detector identifies neutrino interactions by detecting ionization and scintillation light produced by charged particles moving through the detector. Ionization charge is recorded across three wire planes (two induction, one collection) oriented at different angles (with induction planes at  $\pm60^\circ$ relative to the vertical collection plane), and 3 mm wire-spacing. This configuration allows the generation of mm-scale resolution, three-dimensional images of neutrino interactions. Scintillation light collected by the PMTs provides ns-scale timing information, essential for identifying neutrino interactions that are synchronized with the BNB, and for rejecting cosmic-ray backgrounds.

In MicroBooNE, the LArSoft framework \cite{Snider:2017wjd} provides tools for simulating the neutrino flux, neutrino interactions, particle propagation, and detector response including ionization and scintillation processes as well as reconstruction of neutrino events. The BNB neutrino flux through the MicroBooNE detector is simulated using the MiniBooNE flux model \cite{PhysRevD.79.072002} adapted for MicroBooNE's location along the beamline. GENIE v3.0.6 (G18\_10a\_02\_11a)  \cite{Andreopoulos:2009rq, PhysRevD.104.072009} is the event generator used to simulate neutrino-argon interactions, including both single and associated kaon production. In GENIE, associated kaons are produced via individual resonances simulated by the Rein-Sehgal model \cite{REIN198179}. Kaons produced via deep inelastic scattering are simulated by the Koba-Nielsen-Olesen (KNO) scaling parameterization \cite{KOBA1972317}. For higher energies (hadronic invariant mass \textit{W} $>$ 2.3 GeV), the AGKY model \cite{Yang_2009} and PYTHIA6 \cite{Sj_strand_2006} are used. The hadronization models are tuned to match strange particle production ($\Lambda$ and $K^{0}_S$) observed in experiments such as the Big European Bubble Chamber 
\cite{Jones:1992bm, ALLASIA19831, WA59:1991cop, BOSETTI198229} and the Fermilab bubble chamber \cite{PhysRevD.34.1251, PhysRevD.50.6691}. The single kaon production model \cite{skmodel} is implemented in GENIE, generating events in the channels $\nu_{l} + n \rightarrow l^- + n + K^+$, $\nu_{l} + p \rightarrow l^- + p + K^+$, and $\nu_{l} + n \rightarrow l^- + p + K^0$. The region of validity assumed for the model is for $E_{\nu} \leq$ 2 GeV \cite{skmodel, skmodel2}, while the GENIE authors note that the model can be applied to higher energies as well \cite{geniesingle}. 

Particle trajectories and interactions within the detector are simulated using GEANT4 v4\_10\_3\_p03c using the QGSP\_BERT hadron physics list \cite{AGOSTINELLI2003250, 1610988, ALLISON2016186}, which models particle propagation in the detector. Simulated neutrino interactions are overlayed with data events collected using an unbiased trigger that operates in anti-coincidence with the beam, enabling data-driven modeling of cosmic ray and detector noise. 
Both simulation and data reconstruction use the Pandora framework \cite{pandora}, which uses a multi-algorithm approach to pattern recognition and event reconstruction in LArTPCs. Pandora creates clusters from wire hits, identifies the neutrino vertex, matches clusters across wire planes to form 3D objects, classifies reconstructed particles as tracks or showers, and establishes particle hierarchies by identifying parent-daughter relationships among particles.

Before performing a dedicated selection for $K^{+}$ production signal events, a pre-selection is applied as follows: (1) events must pass a CC inclusive filter which isolates CC neutrino interactions by identifying the presence of an outgoing muon from the vertex of a neutrino interaction \cite{MicroBooNE:2019nio}, (2) all reconstructed tracks in an event must originate within a 2.4 m × 2.1 m × 9.8 m fiducial volume centered inside the MicroBooNE TPC, and (3) the daughter track (defined as a track connected to the endpoint of one primary signal track candidate) should stop at least 5 cm away from any of the physical edges of the TPC to ensure containment. 

\begin{figure}[htbp]
\centering
\includegraphics[width=1\linewidth]{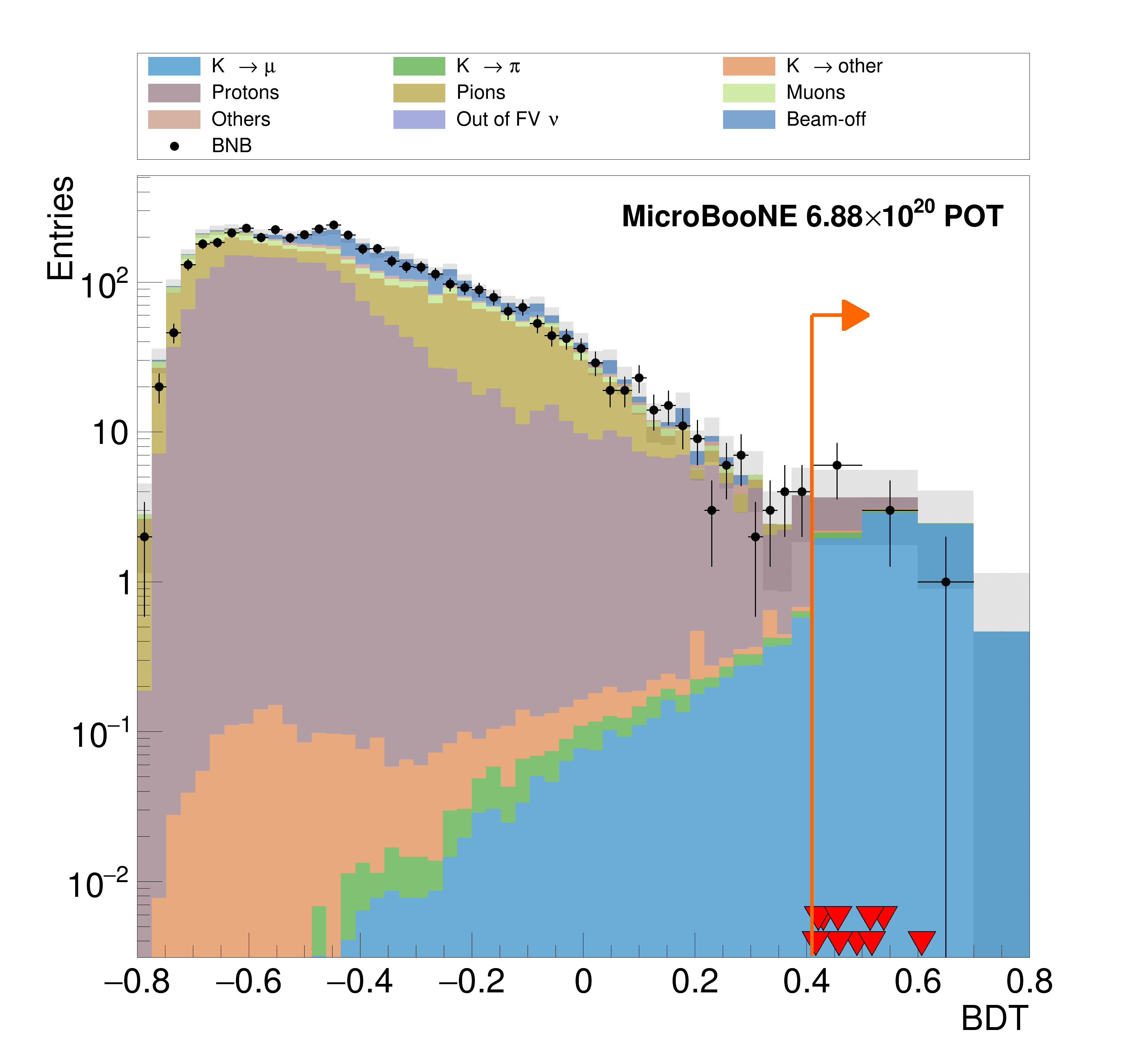}
\caption{BDT classifier score distribution. The black dots represent the MicroBooNE data. The stacked histogram represents the expected BDT distribution from the simulation with its statistical uncertainty (gray band). The orange arrow represents the BDT event selection cut. The red triangles are the $K^{+}$ candidates selected in MicroBooNE data by the BDT.}
\label{fig:bdtscore}
\end{figure}

For the $K^{+}$ event selection, we implemented a boosted decision tree (BDT) classifier \cite{tmvaroot}. The BDT classifier inputs include reconstructed variables such as $\chi^2$ values which compare the expected $dE/dx$ for kaon ($\chi^2_{k}$), proton ($\chi^2_{p}$), pion ($\chi^2_{\pi}$), and muon ($\chi^2_{\mu}$) hypotheses with the observed data for each wire plane, the combined 3-wire-plane $\chi^2$, the daughter track length, and the log-likelihood ratio particle identification \cite{nicpaper2021}. We trained the BDT classifier to identify true  $K^{+}$ that decay to $ \mu^+ \nu_\mu $ or $ \pi^+ \pi^0 $, using Monte Carlo (MC) simulation samples of reconstructed neutrino interactions that meet the preselection criteria described above. To train the BDT classifier, a background sample corresponding to $3.67\times10^{20}$ POT (equivalent to 10\% of the MicroBooNE BNB simulation) was used. For the simulation of the signal, a dedicated high-statistics sample of single and associated $K^{+}$ production was used. We optimized the BDT classifier score selection criteria by maximizing the product of efficiency and purity. Figure \ref{fig:bdtscore} shows the BDT classifier score distribution, with the orange arrow at 0.41 representing the selected BDT score cut. Events with a score higher than the BDT selection are kept as $K^{+}$ candidates, and 10 such $K^{+}$ candidates are identified in the data. 
The total number of background events predicted from neutrino-induced interactions, cosmic-rays recorded during neutrino-triggered beam-spills, and off-beam data is 2.21 events.
The simulation predicts a selection efficiency of 3.95\%, and an associated selection purity of 79\%. The primary background category for $K^+$ candidates consists of protons generated at the neutrino vertex that interact with an argon nucleus, producing secondary particles at the proton endpoint that could mimic a $K^+$ decay product. This background category comprises 2.08 out of 2.21 predicted background events.

\begin{figure}[htbp]
\centering
\includegraphics[width=1\linewidth]{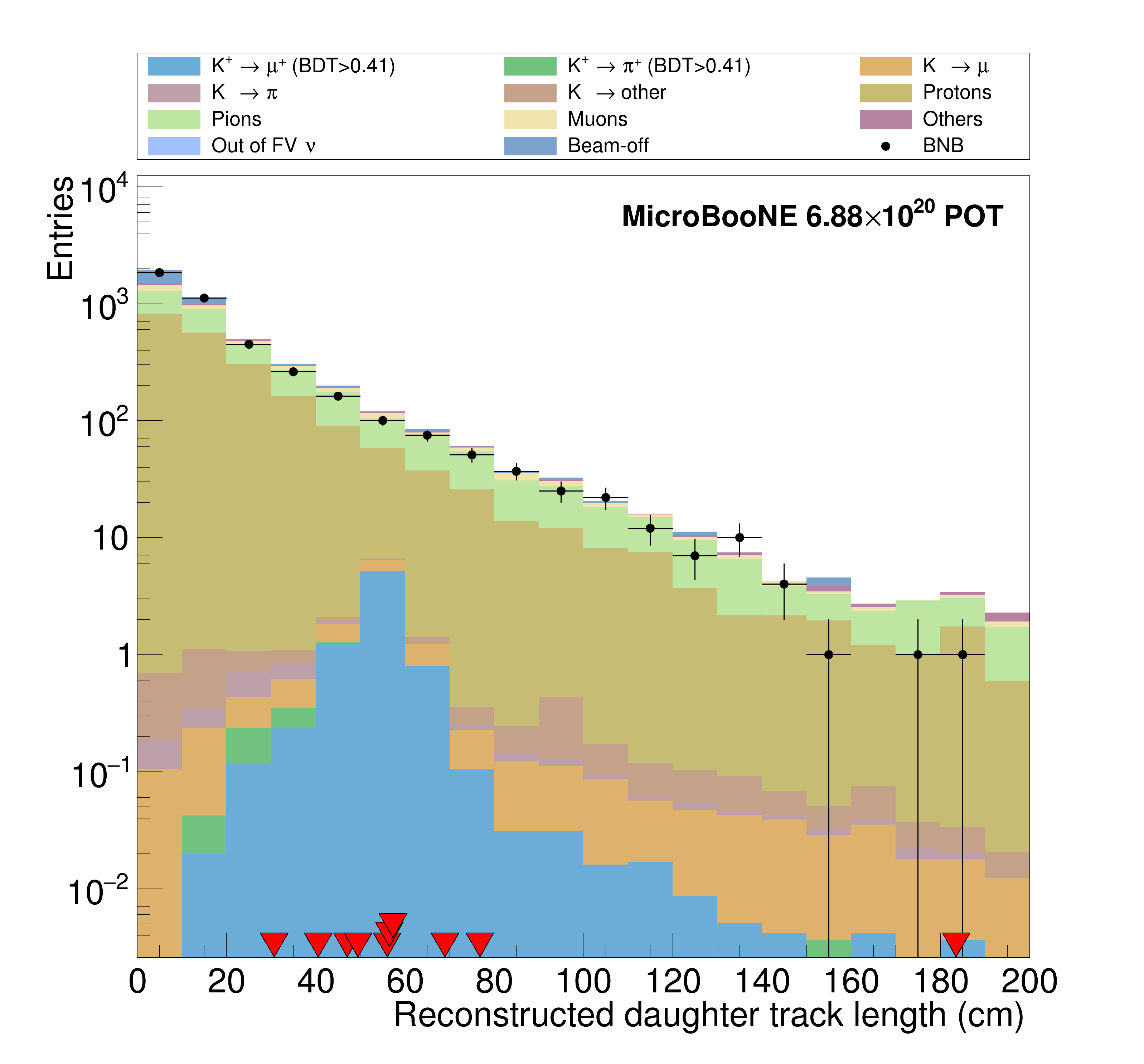}
\caption{Reconstructed daughter track length distribution. The stacked histogram shows the MC simulation with its statistical uncertainty (gray band). The black dots represent the MicroBooNE data. The red triangles are the $K^{+}$ candidates selected in MicroBooNE data by the BDT, where three data points lie close to 60 cm, nearly overlapping.}
\label{fig:kpikmuDG}
\end{figure}

Figure \ref{fig:kpikmuDG} shows the reconstructed daughter track length distribution of the $K^{+}$ candidates selected in MicroBooNE data, including the event shown in Fig. \ref{fig:evd}. The stacked histogram shows the expected daughter track length distribution from the MicroBooNE BNB simulation. Due to the two-body decay nature of the $K^+$ branching ratios analyzed in this letter, the resulting monoenergetic $\mu^+$ ($\pi^+$)
provides a powerful discriminant, as it leads to a fixed track length that can be measured with high accuracy and is not present in background events. The expected track length for $\mu^+$ ($\pi^+$) accumulates around 53 cm (30 cm).
The $K^{+}$ candidates selected by the BDT selection in Fig. \ref{fig:kpikmuDG} (red triangles) exhibit a pile-up in the region compatible with $K^{+}$ decay. 

To evaluate the accuracy of the background modeling, two sideband regions were defined using a 2-dimensional distribution of the BDT classifier score versus daughter track length. A far sideband region is defined to include tracks with low BDT classifier scores (BDT $<$ 0) and track lengths less than 40 cm. A near sideband region selects tracks with BDT classifier scores close to the signal region (BDT $<$ 0.41) and across all track length ranges, but excludes tracks in the far sideband region. The data/MC comparisons of the near and far sidebands show agreement for reconstructed kinematic variables such as $K^{+}$ candidate track length, $\phi$ (azimuthal angle around the beam direction), and $\theta$ (angle of the track with respect to the beam direction). For the $K^+$ candidate track length data/MC comparison, 
$\chi^2/\text{ndf}$ values (including statistical uncertainties) of 16.87/24 and 17.93/24 were obtained for the near and far sidebands. The agreement with the data suggests that the simulation is able to adequately model the background in the cross-section extraction. More information about the sideband comparisons is provided in the Supplemental Material \cite{suppmat}.

This analysis presents the measurement of the single-bin flux-integrated cross section for $\nu_{\mu} + \text{Ar} \rightarrow \mu^{-} + K^{+} + X$ interactions, where muon neutrinos interact with Ar and produce one $\mu^{-}$, one $ K^{+}$ and $X$ (any number of other hadrons in the final state) on argon nuclei. 
The cross section ($\sigma$) was found to be 7.93 $\pm$ 3.22 (stat.) $\pm$ 2.83 (syst.) $10^{-42} \text{cm}^2/{\rm nucleon}$, and it was calculated using the following expression:
\begin{align}
\sigma = \frac{N - B}{\varepsilon \times N_{\text{target}} \times \Phi_{\nu_{\mu}}}~\text{,}
\end{align}
where $N$ represents the number of selected signal events (10),
$B$ is the number of expected background events (2.21), 
$\varepsilon$ is the selection efficiency ($3.95\%$),
$N_{\text{target}}$ is the number of target nucleons ($4.13 \times 10^{31}$), and
$\Phi_{\nu_{\mu}}$ stands for the integrated BNB $\nu_{\mu}$ flux ($6.02 \times 10^{11} \nu/\text{cm}^2$). The selection efficiency accounts for all $K^+$ decays within the fiducial volume.

The systematic uncertainties in the measured cross section are evaluated by reweighting and generating simulated events to account for each source of uncertainty \cite{MicroBooNE:2021ccs}. For the detector-related uncertainties, a dedicated simulation is used to model the variation in detector responses \cite{Abratenko_2022, Calcutt_2021}.
In contrast, uncertainties from neutrino interactions and BNB flux modeling are treated with a reweighting technique. 
To evaluate the uncertainty associated with kaon reinteractions within the detector, we implement a methodology similar to that presented by MINERvA \cite{PhysRevD.94.012002}. For each selected $K^+$ candidate track, a weight $W_{\text{inel}}$ is calculated using the following expression to estimate the kaon inelastic reinteraction uncertainty:
\begin{equation}
W_{\text{inel}} = \frac{1 - e^{-\rho x \sigma^{\text{tot}}_{\text{var}}}}{1 - e^{-\rho x \sigma^{\text{tot}}_{\text{geant}}}} \times \frac{\sigma^{\text{inel}}_{\text{var}}}{\sigma^{\text{tot}}_{\text{var}}} \times \frac{\sigma^{\text{tot}}_{\text{geant}}}{\sigma^{\text{inel}}_{\text{geant}}}~\text{,}
\label{eq:inel}
\end{equation}
where $\rho$ is the density of the material, $x$ is the track length, $\sigma^{\text{tot}}_{\text{geant}}$ and $\sigma^{\text{inel}}_{\text{geant}}$ are the $K^+$ total and inelastic GEANT4 cross sections, and $\sigma^{\text{tot}}_{\text{var}}$ and 
$\sigma^{\text{inel}}_{\text{var}}$ are the $K^+$ total and inelastic cross sections estimated by applying a variation of $\pm$40\% to the $\sigma^{\text{inel}}_{\text{geant}}$. 
The 40\% variation is motivated by observed disagreement between experimental measurements of $K^+ - \text{Ca}$ interactions \cite{PhysRevC.55.1304} and the predictions from the GEANT4 interaction model shown in the supplemental material \cite{suppmat}.

The leading source of systematic uncertainty comes from the detector response (30.3\%), with a major contribution from the modeling of ion recombination of electrons with argon. This analysis relies on identifying stopping $K^+$ and $\mu^+$ ($\pi^+$) through their Bragg peaks, which are characterized by a sharp rise in energy deposition near the end of the particle track. These effects occur in a highly ionizing regime, where ion recombination effects are most pronounced. As the charge profile is impacted by recombination, any mismodeling can distort the Bragg peak shape and affect particle identification. 
The modeling of neutrino interactions, the BNB neutrino flux, and particle reinteractions in the detector, including kaon inelastic interactions, lead to systematic uncertainties of 3.7\%, 11.7\%, and 10.5\% respectively.
The neutrino interaction modeling uncertainty on the cross-section measurement enters through the modeling of the subtracted background and the signal efficiency prediction. The neutrino interaction modeling uncertainty was evaluated by varying the GENIE parameters within their nominal 1$\sigma$ ranges \cite{MicroBooNE:2021ccs}. These variations were propagated through the full simulation to assess their impact on the measured kaon production cross section uncertainty budget. The main contributions to the neutrino interaction modeling arise from CC resonance production (axial and vector masses). The resulting uncertainty shows limited impact on the selection efficiency, as the selection depends mainly on topological and calorimetric features rather than interaction kinematics which are more tied to detailed cross-section modeling.
The effect of statistics in the estimation of the selection efficiency (6.1\%), the number of target nuclei $N_{\text{target}}$ (1\%), MC statistical uncertainties on the background estimation (7.4\%), and POT exposure (2\%) are the remaining sources of uncertainty. The overall magnitude of the uncertainty (statistical and systematic) on our measurement is 54.1\%, where the major contribution is due to the statistical uncertainty, with a reported value of 40.6\%. Table \ref{tab:cross_sections} presents the measured cross section of the  $\nu_{\mu} + \text{Ar} \rightarrow \mu^{-} + K^{+} + X$ interactions and the predictions obtained from various neutrino event generators with the NUISANCE framework \cite{Stowell:2016jfr}.

\begin{table}[htbp]
    \centering
    \begin{tabular}{c|c}
        \textbf{Generator} & \textbf{cross section ($10^{-42} \text{cm}^2/\text{nucleon}$)} \\ \hline
        GENIE v2.12.10 & 8.67  \\ 
        GENIE v3.00.06 & 8.42  \\
        GENIE v3.4.0 (AR23) & 9.85 \\
        GiBUU 2025, patch 1 & 6.52  \\ 
        NEUT 5.4.0.1 & 9.71  \\ 
        NuWro 19.02.1 & 10.87  \\ \hline
        \textbf{MicroBooNE Data} &  \textbf{7.93 $\pm$ 3.22 (stat.) $\pm$ 2.83 (syst.)}\\ \hline
    \end{tabular}
    \caption{The $\nu_{\mu} + \text{Ar} \rightarrow \mu^{-} + K^{+} + X$ cross section extracted from different neutrino event generators compared with the cross section extracted from data.}
    \label{tab:cross_sections}
\end{table}

The flux-integrated cross-section measurement reported in this paper indicates consistency with the predictions from different neutrino event generators, although the large uncertainties prevent a stringent test of these models. 
The cross sections predicted by NuWro \cite{GOLAN2012499} and GiBUU 2025, patch 1 \cite{gibuu} show larger deviations from the data, but remain within the 1$\sigma$ uncertainty of the measurement.

In conclusion, this letter presents the first measurement of the $\nu_{\mu} + \text{Ar} \rightarrow \mu^{-} + K^{+} + X$ cross section on argon, using the MicroBooNE dataset collected from the BNB flux.
The results reported here represent a significant step forward for future studies aiming to refine background estimates for proton decay searches, particularly within the framework of GUTs and SUSY. To advance this analysis in future LArTPC measurements, it is necessary to increase statistics, reduce large detector systematics (especially those related to the recombination of electrons with argon nuclei), refine reconstruction algorithms for short tracks, and improve the identification of kaon tracks that undergo interactions. Experiments such as SBND \cite{sbndcollaboration2025shortbaselineneardetectorfermilab}, ICARUS \cite{AMERIO2004329}, and the DUNE near detector will allow precision measurements of neutrino-induced $K^+$ production on argon, including differential cross-section measurements for the first time ever. 
Furthermore, the uncertainty related to kaon reinteractions on argon nuclei could be refined by using data from LArIAT \cite{osti_1489387}, and the DUNE prototypes at CERN \cite{PhysRevD.110.092011}. More statistics will come from analyzing MicroBooNE's full dataset with a total of 1.2 $\times 10^{21}$ POT. The future DUNE near detector data could also provide more accurate measurements of CC neutrino-induced $K^{+}$ production on argon nuclei enabling improved comparisons among the neutrino generator models. These high-precision measurements will enhance our understanding of rare neutrino interactions and final state interaction modeling, and will improve the background predictions for future nucleon decay searches at DUNE, which currently rely on model predictions.

This document was prepared by the MicroBooNE collaboration using the resources of the Fermi National Accelerator Laboratory (Fermilab), a U.S. Department of Energy, Office of Science, Office of High Energy Physics HEP User Facility. Fermilab is managed by Fermi Forward Discovery Group, LLC, acting under Contract No. 89243024CSC000002. MicroBooNE is supported by the following: the U.S. Department of Energy, Office of Science, Offices of High Energy Physics and Nuclear Physics; the U.S. National Science Foundation; the Swiss National Science Foundation; the Science and Technology Facilities Council (STFC), part of the United Kingdom Research and Innovation; the Royal Society (United Kingdom); the UK Research and Innovation (UKRI) Future Leaders Fellowship; and the NSF AI Institute for Artificial Intelligence and Fundamental Interactions. Additional support for the laser calibration system and cosmic ray tagger was provided by the Albert Einstein Center for Fundamental Physics, Bern, Switzerland. We also acknowledge the contributions of technical and scientific staff to the design, construction, and operation of the MicroBooNE detector as well as the contributions of past collaborators to the development of MicroBooNE analyses, without whom this work would not have been possible. For the purpose of open access, the authors have applied a Creative Commons Attribution (CC BY) public copyright license to any Author Accepted Manuscript version arising from this submission.

\bibliography{main} 

\end{document}